# The reentrant condensation of polyelectrolytes induced by diluted multivalent salts: A mean-field level revisiting

Huaisong Yong[1,2,3]*, Bilin Zhuang[4]*, and Sissi de Beer[1]*

[1]Department of Molecules & Materials, MESA+ Institute, University of Twente, 7500 AE Enschede, The Netherlands

[2]Institute Theory of Polymers, Leibniz-Institut für Polymerforschung Dresden e.V., D-01069 Dresden, Germany

[3]School of New Energy and Materials, Southwest Petroleum University, 610500 Chengdu, China

[4]Department of Chemistry, Harvey Mudd College, 301 Platt Blvd, Claremont, CA 91711, USA

*Correspondences: h.yong@utwente.nl (H.Y.); bzhuang@hmc.edu (B.Z.); s.j.a.debeer@utwente.nl (S. de B.)

## Abstract

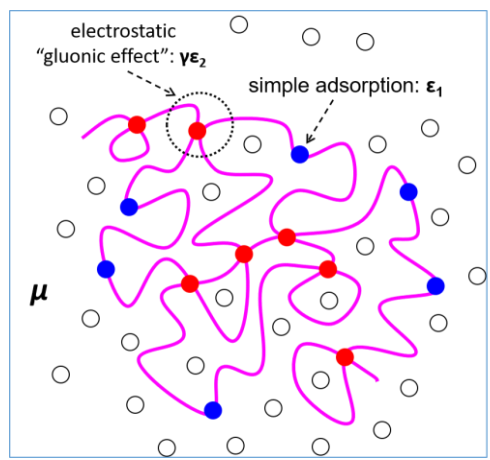
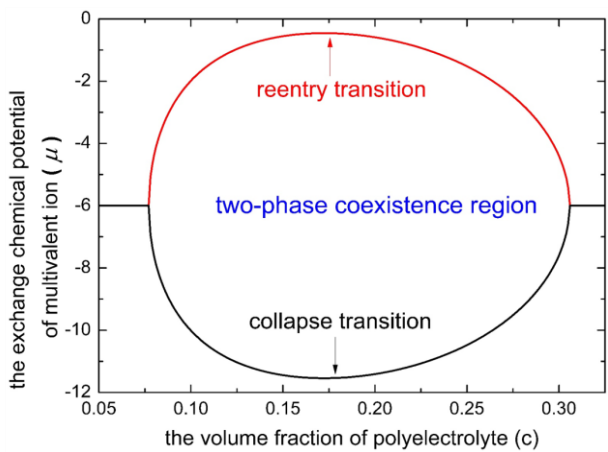

We study the reentrant condensation of polyelectrolytes induced by diluted small multivalent salts, whose phase-transition mechanism remains poorly understood. Motivated by recent all-atom simulation results reported by the Caltech group on phase behaviors of polyelectrolytes in the presence of multivalent salts (**Macromolecules, 2024, 57, 1941–1949; Langmuir, 2024, 40, 6212–6219**), in this work we construct a simple but effective mean-field model which can rationalize the essential features of the reentrant condensation including the phase diagram of polyelectrolyte. For the first time, we separate the electrostatic effect into electrostatic gluonic effect due to sharing multivalent salt ions between ionic monomers, and the non-associative pair-wise electrostatic effect by correlation of all ions. This approach allows us to uncover that the electrostatic gluonic effect rather than other effects dominates the reentrant phase transition of polyelectrolyte induced by diluted multivalent salts. The model unveils that the strong adsorption between the ionic monomers and multivalent ions can be at the origin of the peculiar phenomenon that rather low concentrations of multivalent salts trigger both collapse and reentry transitions. The analytical solution of the model indicates that a minimum coupling energy due to sharing multivalent salt ions between ionic monomers is essential for a phase transition to occur, which for the first time explains the enigmatic observation that polyelectrolytes can only show phase transition in a dilute solution of salts with selective multivalency. Our analytical calculations also show that the incompatibility of the uncharged moieties of the polyelectrolytes with water is critical to regulate phase behaviors of polyelectrolytes in aqueous solutions. This is in agreement with recent experimental investigations on the solution properties of amphiphilic proteins. The obtained results will contribute to the understanding of liquid-liquid phase separation in biological systems where multivalent ions bound to bio-polyelectrolytes play an essential role.

## 1. Introduction

Polyelectrolytes such as proteins mixed with multivalent salts, rich phenomena of phase transitions can occur by variations in salt compositions, temperature and polymer concentration, which have attracted growing attention in the past decades **[1-4]**. A particular scenario is the reentrant condensation of proteins in the aqueous solution of small multivalent salts, as sketched in **Figure 1**, where the collapse transition occurs when a rather small concentration of multivalent counterions is introduced (such as tripolyphosphate, $Fe^{3+}$ and $Y^{3+}$), and the reentry transition of the precipitated proteins can often happen when a minor excess of the multivalent counterions is further added **[5-8]**. A remarkable aspect of this reentrant transition is that: The concentration of overall added multivalent salt is quite low (usually in the order of 5 mmol/L) when the reentrant condensation occurs **[5, 8]**, which implies that the effect of electrostatic screening actually plays a trivial role, since for this case the Debye length of electrostatic screening is comparable with the size of an isolated polyelectrolyte chain or even larger **[9-11]**

Clarifying the function of diluted multivalent salts in the reentrant condensation of polyelectrolytes is pivotal for a better understanding of liquid-liquid phase separation in biological systems if multivalent ions bound to bio-polyelectrolytes (such as RNAs **[12]** and proteins **[13]**) plays an essential role **[1-4, 14-16]**. However, by classical polyelectrolyte theories **[17-21]**, it remains a challenge to understand/analyze the reentrant condensation of polyelectrolytes in the presence of diluted multivalent salts. For example, the well-known mechanism of counterion condensation **[17-21]** can lead to a collapse transition of polyelectrolytes but not a reentry transition when salt concentration increases, which cannot explain the re-solubilization of some amphiphilic proteins when a minor excess of the multivalent counterions is further introduced **[5, 8]**. We note that electric dipoles can be formed between ionic monomers and counterions under certain conditions, and the electric dipole-dipole attraction can be partially attributed to the collapse transition of polyelectrolytes **[9, 22]**. But, the strength of an electric dipole-dipole attraction is generally enhanced with the addition of salt, which cannot induce the reentry transition of polyelectrolyte in diluted salt solutions as illustrated by **Figure 1**. It is also worthy of pointing out that recent theories **[23, 24]** for the coacervation of polyanions with polycations may somehow explain the collapse transition of polyelectrolytes in solutions of small multivalent salts. Nevertheless, because the physical foundation of those coacervation theories **[23, 24]** mostly relies on the fact that the release of small counterions from polyelectrolyte chains entropically drives the collapse transition and the reentry transition cannot be expected without massive addition of small salts **[25]**. Thus, they are not suitable to explain the reentry transition of polyelectrolytes in diluted salt solutions **[5, 8]**.

To understand the reentrant condensation of polyelectrolytes in dilute solutions of multivalent salts, an alternative way to explain can be as follows. We consider an adsorption process where the ionic monomers are the substrate for ions. When there is a higher affinity of ionic monomer with multivalent ions compared to mono-valent ions, for example, in the system of polyacrylic acid/sodium ion/rare-earth metal ions **[26]**, the multivalent ions prefer to adsorb on the ionic monomers and form coordination complexes. Notwithstanding, if we solely take into account pairwise-like interactions for preferential adsorption between multivalent ions and ionic monomers, linear polymers are 1D substrates for ions. From a theoretical consideration, this situation can be taken into account in formal analogy with the 1D-Ising model **[27, 28]**: Therefore, one should not expect a phase transition merely according to the simple adsorption effect.

Thus, it is necessary to further account for an additional attraction between multivalent ions and ionic monomers. A quite natural way to realize the attraction interaction is the share of a multivalent ion by several ionic monomers, i.e., forming "physical crosslinks" between ionic monomers by sharing multivalent ions as sketched in **Figure 2**. Such "crosslinking attraction" also enhances the adsorption of multivalent ions by increasing the adsorption energy, which was corroborated in detail by recent all-atom simulation studies of the Caltech group **[29, 30]** on the phase behaviors of poly(acrylic acid) and sodium polyacrylate in aqueous solutions of calcium salts. We note that similar electrostatic "crosslinking attraction" effect even exists between multivalent ions and polar atoms of uncharged proteins **[31]**. To reach more binding with multivalent ions, a collapsed conformation can be preferred when the attraction strength is strong enough, even if the entropy of chain conformations is reduced. This approach introduces a coupling between adsorption and attraction and in turn leads to an effective nonlinear attraction interaction between ionic monomers beyond the pairwise-like interaction, which cannot be addressed by classical theories **[17-21]** such as the "double screening theory" **[19, 32]** that mainly consider the pairwise electrostatic interaction from the correlations of all ions.

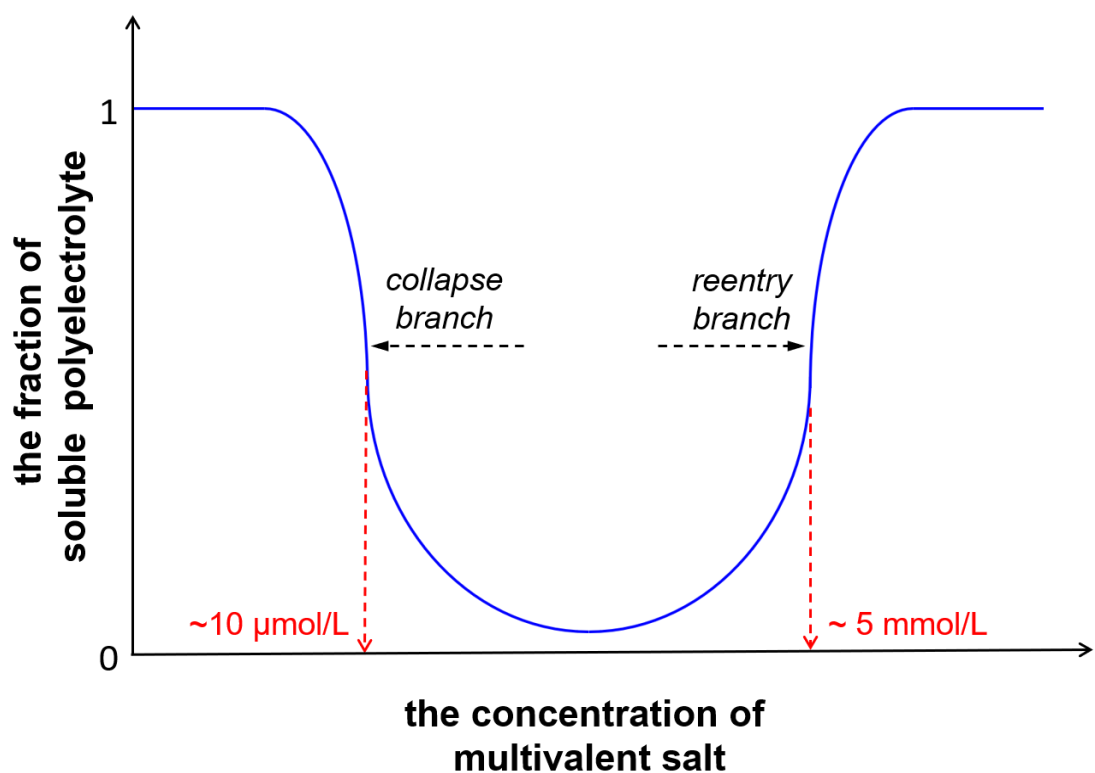

**Figure 1.** A sketch description of the reentrant condensation of polyelectrolytes induced by diluted multivalent salts. In the figure, the typical values of multivalent salt concentration for collapse and reentry transitions are quoted from **refs. [5, 8]**. Note that the reentrant condensation is not necessary to be symmetric with respect to multivalent salt concentrations.

A point worthy of note is that: Although previous theoretical formalisms **[33-45]** for the concept of "physical crosslinks" such as "ion bridges", "complexations" or "short-range attraction" have promoted significantly the understanding of phase behaviors of polyelectrolytes in the presence of multivalent salts; however, they failed to address the experimental fact that polyelectrolytes can only show phase transition in the dilute solution of salts with selective multivalency **[46-50]**. For example, bovine serum albumin proteins show phase transition in aqueous solutions of tri-valent and tetra-valent salts but do not show phase transition in aqueous solutions of di-valent salts **[47]**. In addition, most of previous theoretical formalisms **[33-45]** overlooked the influence of non-electrostatic interactions such as hydrophobic interaction, which indeed plays an important role in regulating phase transition of polyelectrolyte as indicated by recent experimental investigations on solution properties of proteins **[51-55]**.

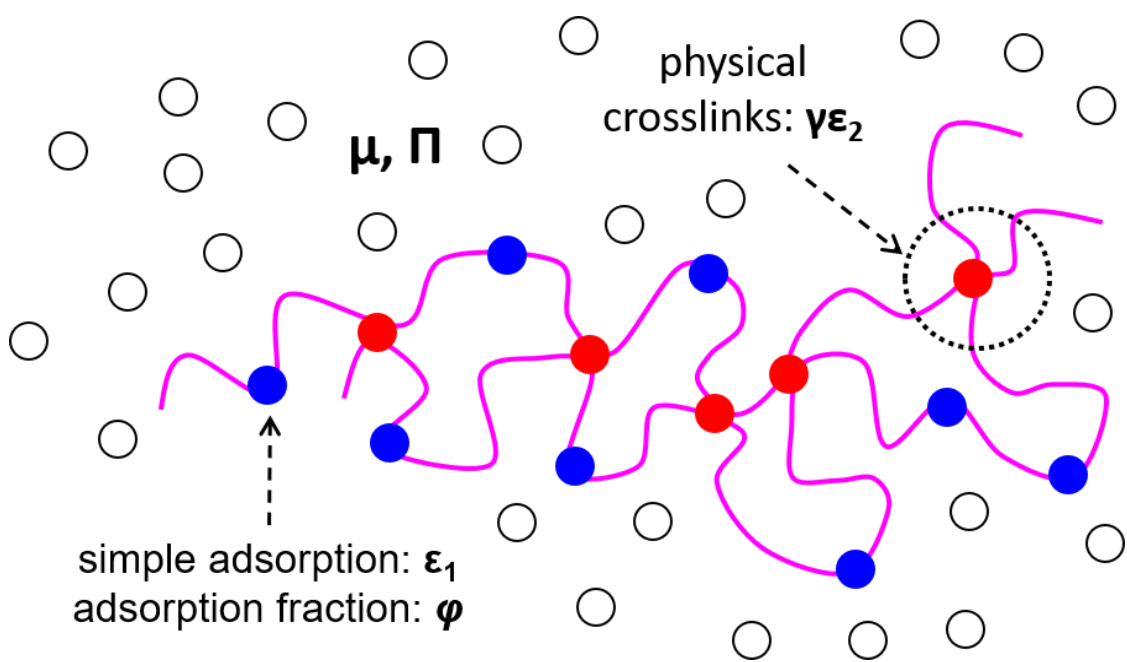

**Figure 2.** A sketch description of preferential adsorption of multivalent ions by ionic monomers (filled blue circles), and forming "physical crosslinks" between ionic monomers by sharing multivalent ions (filled red circles). Here the role of the multivalent ion is just like a glue to bind different ionized monomers. In the figure, the pink lines represent polymer chains and open black circles represent background ions in polyelectrolyte solutions.

In parallel, it is worth of pointing out that a quite similar situation exists in the research of the cononsolvency effect of polymers (a typical reentrant condensation of polymers), where the binary mixtures of two good solvents can be poor solvents of the same polymer **[56]**. With analogy, we note that the concept of cosolute-assisted "physical crosslinking" effect proposed by Sommer **[57, 58]** to explain cononsolvency effect, which can be extended to rationalize the current level of research **[1-4]** on the reentrant condensation of polyelectrolytes in dilute solutions of small multivalent salts.

Bearing these points in mind, in this work, by viewing the multivalent ion as a kind of charged "gluonic cosolute" **[57, 58]**, we constructed a simple physical model and found that the strong adsorption between the ionic monomer and multivalent ion can contribute to the peculiar phenomenon of rather low concentrations of multivalent salts triggering both collapse and reentry transitions. The analytical solution of the model indicates that a minimum coupling energy due to sharing multivalent salt ions between ionic monomers is essential for phase transition to occur, which explains the puzzling observation that polyelectrolytes can only show a phase transition in dilute solutions of salts with selective multivalency. Our calculations also show that the incompatibility of the uncharged moieties of polyelectrolyte with water is critical to regulate the phase behaviors of polyelectrolyte in aqueous solutions.

In the remainder of this article, the physical model for the polyelectrolyte solution in terms of the free energy will be constructed in **Section 2**. Its analytical solution will be considered in detail in **Section 3** and **Section 4**, where we will outline some general consequences of the model and a simplified phase diagram of the polyelectrolyte solution will be discussed in these sections. Finally, the applicability of the model will be discussed in **Section 5** with concluding remarks.

## 2. The construction of Gibbs free energy and its physics foundation

For simplicity but without losing generality, as sketched in **Figure 2**, we consider flexible polyelectrolyte with monovalent ionic monomers and monovalent counterions when no salt is introduced. We denote $N$ as the number of monomers in a polyelectrolyte chain and $a$ as the size of each monomer along the direction of the polymer backbone. The charged monomers are distributed randomly on the polyelectrolyte chain and their fraction is denoted by $p$. Then the average distance between neighboring charged monomers is $a/p$. There is a saturated charge density for polyelectrolyte chains due to the repulsion correlation effect in the dissociation of neighboring ionic monomers **[10]**. The saturated charge density is determined by the condition that polyelectrolyte does not undergo counterion condensation **[17-21]**:

$$\frac{e^2}{4\pi\varepsilon_w k_B T}\left(\frac{p}{a}\right) < 1. \tag{1}$$

Here, $e$ is the electron charge, $\varepsilon_w$ is the absolute dielectric permittivity of water, $k_B$ is the Boltzmann constant, and $T$ is thermodynamic temperature. Because the Bjerrum length $l_B = e^2/(4\pi\varepsilon_w k_B T)$ is about 0.74 nm for water and $a$ is about 0.25 nm for vinyl polymers (the length of two carbon-carbon single bonds), we see that $p < a / l_B \approx 1/3$, which is usually satisfied by amphiphilic polyelectrolyte such as proteins **[59, 60]**. On account for this consideration and to simplify our discussion, in this study we focus on polyelectrolyte whose charge density is under the saturated charge density, i.e., the case of $p < a / l_B \approx 1/3$.

Here and in the following, let's consider the free energy per unit of volume for an incompressible system if not otherwise noted specifically. The volume unit is given by the size of the solvent molecules in the spirit of the Flory–Huggins lattice model. For simplicity but without losing generality, we assume that the sizes of the ionized monomers, counterions, salt ions and solvent molecules are the same. The added multivalent salt has the chemical formula $XZ_x$, where X is the multivalent cation with the valence $+x$, Z is the monovalent anion. The overall volume fraction of monovalent ionic monomers and neutral monomers is denoted by $c$, then the volume fraction of monovalent counterions is $pc$. We denote the volume fraction of multivalent cations X as $c_x$, then the volume fraction of monovalent anion Z is $xc_x$.

The Gibbs free energy in general consists of many terms. The most common term is the contribution from the mixing of polymer with solvent and added multivalent salt, $G_{sol}$. Within the mean-field approximation, the free energy of $G_{sol}$ is given by

$$G_{sol} = \frac{c}{N}\ln(c) + pc\ln(pc) + c_x \ln(c_x) + xc_x \ln(xc_x) \\ + \left[1-(1+p)c-(1+x)c_x\right]\ln\left[1-(1+p)c-(1+x)c_x\right] + \Pi \quad . \tag{2}$$

As sketched in **Figure 2**, here we separate the system into two parts; the polymer chains with their enclosed solvents and small ions, and the bulk without polymer chains. The bulk is just described by its osmotic pressure ($\Pi$) acting on the first part. If the volume of the polymer coils changes, then there is the mechanical work against the external pressures involved in the free energy change. This will become important when we minimize the free energy with respect to the monomer concentration ($c$). Here and in the following, we consider energies in units of $k_B T$ if not otherwise noted specifically.

Since the exchange of counterions only occurs on the ionized monomers, the adsorption free energy per unit of volume owing to the mixing of multivalent and monovalent cations on the polymer chains is therefore given by $G_{ads}$,

$$\frac{G_{ads}}{pc} = \varphi \ln(\varphi) + (1-\varphi)\ln(1-\varphi) - \mu\varphi - \varepsilon_1\varphi . \qquad (3)$$

Here, $\varphi$ is the fraction of ionic monomers occupied preferentially by multivalent cations. $\varepsilon_1$ denotes the preferential-adsorption strength of one multivalent cation with respect to the polyelectrolyte, which stems primarily from ionic-bond interactions or electric dipole interactions and is often not small (on the order of about 5 $k_BT$ for the strength of an ionic bond in water at low salt concentration **[61]** and in the order of about 10 $k_BT$ for the strength of electric dipole interaction **[62]**). $\mu$ represents the chemical potential of exchanging a monovalent cation for a multivalent cation on the polymer chains, which scales as $\mu \sim \ln(c_x)$ if $c_x$ is very small.

We consider the associative electrostatic attraction, $G_{attr}$, between ionized monomers caused by forming a bridge due to multivalent cations, where the bridge is a kind of short-range attractive interaction as proven numerically by the Gaussian renormalized fluctuation theory recently **[36]**. As sketched in **Figure 2**, here the role of the multivalent cation is just like a glue to bind different ionized monomers, which is in analogy to the concept of "gluonic cosolute" proposed by Sommer **[57, 58]** to explain cononsolcency effect **[56]**. Statistically, the term of $G_{attr}$ is proportional to the average probability that a given ionized monomer ($\varphi$ or $1 - \varphi$ states) is in contact with an ionized monomer of the other state ($1 - \varphi$ or $\varphi$). Here, the state of the monomer is defined by either having adsorbed a multivalent cation (with probability $\varphi$) or being empty (only surrounded by solvent and monovalent cation, with probability $1 - \varphi$). The strength of this additional attraction (electrostatic "gluonic effect") is given by $\gamma\varepsilon_2$, which takes into account the bridging efficiency of the molecular matching **[46]** between the ionized monomer and the multivalent ion by the numerical coefficient $\gamma \neq 0$. Since the electrostatic "gluonic effect" only occurs among ionized monomers, the attraction energy owing to the electrostatic "gluonic effect" is therefore given by the

mean-field approximation

$$\frac{G_{attr}}{pc} = -2\gamma\varepsilon_2\varphi(1-\varphi)(pc). \tag{4}$$

The ionic-bond features of adsorption and attraction imply that $\varepsilon_1 \approx \varepsilon_2$ due to the chelation of multivalent ions by ionic monomers. A physical boundary condition is embedded in $G_{attr}$, that is both $\varphi \to 0$ and $\varphi \to 1$, leading to the vanishing of $G_{attr} \to 0$ since this condition corresponds to the fact of reentrant transition that polyelectrolyte must be miscible within aqueous solutions with low and high concentrations of multivalent salts respectively. We note that another physical boundary condition can be embedded in $G_{attr}$ by $\gamma = (x-1)\gamma_0$ with $\gamma_0 \neq 0$, that is monovalent $+x = 1$ leading to the vanishing of $G_{attr} \to 0$ since this condition corresponds to the fact that the dilute solution of monovalent salts cannot lead to the phase transition of polyelectrolyte.

A basic characteristic of the statistical formulation of **Equation(4)** is that it shows the number of bridges per ionic monomer first increases then declines with the amount of adsorbed multivalent ions on polyelectrolyte chains ($\varphi$), i.e., $\propto 2\varphi(1-\varphi)(pc)$, which was confirmed qualitatively by recent all-atom simulation results reported by the Caltech group **[29, 30]**. Another characteristic of **Equation(4)** is that it implies the strength of "gluonic effect" ($\gamma\varepsilon_2$) declines with increasing temperature and in turn induces a weaker phase transition, which aligns with recent fluorescence experiments on phase behaviors of protein solutions in the presence of multivalent salts **[63, 64]**. The preferential adsorption of multivalent ions on polyelectrolyte chains with the physical crosslinking of ionic monomers inevitably leads to overcharging and charge inversion **[65]** of polyelectrolyte chains at certain concentrations of multivalent salts, which implies that the free energy of polyelectrolyte solution must have minimum(s) or symmetry breaking(s) around these salt concentrations **[29, 30, 36, 66, 67]**. We note that this fact is qualitatively reflected by the statistical construction of **Equation(4)** with a minimum and a symmetry breaking at $\varphi = 1/2$. Because of this feature, the fraction of polyelectrolyte chains adsorbed preferentially by multivalent ions ($\varphi$) is also the order parameter for the phase transition in our model.

Besides the electrostatic gluonic energy considered by **Equation(4)**, the free energy of non-associative pairwise electrostatic interaction due to correlations of all ions in the low-salt limit is approximated by the classical "double screening theory" **[19, 32]** as $G_{DS}$,

$$G_{DS} = -\frac{(\kappa a)^3}{8\pi} + \frac{4}{81}\left(3\pi \frac{l_B}{a}\right)^{\frac{1}{2}} pc^{\frac{3}{2}}, \qquad (5)$$

where the inverse Debye screening length $\kappa$ is given by $(\kappa a)^2 \approx 4\pi(l_B/a)(pc + (x^2+x)c_x)$. The first term of **Equation(5)** stems from the non-associative attraction correlation of all small ionized ions, which is approximated by the Debye–Hückel theory for the isothermal–isobaric ensemble (*NPT* ensemble) **[68]**. The second term is from the fluctuation contribution of ionized polymer chains, which is always positive because of the electrostatic repulsion between ionized monomers **[69]**.

The free energy contribution by the formalism of **Equation(5)** is negative for experimental values of $l_B/a$, which aligns with classical polyelectrolyte theories **[17-21]**. **Equation(5)** is strictly valid when the electrolyte concentration is small, about less than 5 mmol/L **[68]**, which usually is also the regime for the occurrence of the reentrant condensation of polyelectrolyte **[5, 8]**. In salt solutions, salt ions usually coordinate with polymer backbones with their hydration shells **[31, 70]**. It is thus expected that both the electrostatic adsorption and attraction are not permanent and have a dynamic-bond feature, which was confirmed by recent all-atom simulation studies of the Caltech group **[29, 30]** on phase behaviors of poly(acrylic acid) and sodium polyacrylate in aqueous solutions of calcium salts. To simplify analytical calculations but without losing generality to get key physical conclusions for the current research, the effect of charge neutralization has been accordingly ignored in the formulation of the free energy of electrostatic interaction.

The monomers of polyelectrolyte chains are composed of both charged and uncharged moieties. This is also true for charged monomers. The energy of non-electrostatic excluded-volume interactions between the neutral part of monomers and solvent (water) is given by Flory-Huggins formalism $G_{FH}$,

$$G_{FH} = \left[1 - (1+p)c - (1+x)c_x\right]\left[p\varepsilon_{FH,1} + (1-p)\varepsilon_{FH,2}\right]c, \qquad (6)$$

where $\varepsilon_{FH,1}$ is the Flory-Huggins parameters between solvent and charged monomers, and $\varepsilon_{FH,2}$ is the Flory-Huggins parameters between solvent and uncharged monomers. We point that the hierarchy of non-coulomb interaction ($\varepsilon_{FH,1} \leq \varepsilon_{FH,2}$) usually exists between charged and uncharged monomers.

Then the total free energy per volume unit can be written as

$$G(\varphi, c, c_x) = G_{sol} + G_{ads} + G_{attr} + G_{DS} + G_{FH}. \tag{7}$$

The current model is a mean-field formalism under the framework of the isothermal–isobaric ensemble (*NPT* ensemble). We note that the volume of the polyelectrolyte solution is given by $V = N_m a^3/c$ and changes with the monomer concentration $c$, where $N_m$ is the total number of monomers in the polyelectrolyte solution and is fixed in the formalism of *NPT* ensemble. Thus, in order to find the equilibrium state of the polyelectrolyte phase with respect to the bulk solvent phase, it is necessary to minimize the free energy per monomer, i.e., $G(\varphi, c, c_x)/c$ instead of $G(\varphi, c, c_x)$, with respect to the adsorption fraction of multivalent cation $\varphi$, the polymer concentration $c$, and the concentration of multivalent cation $c_x$.

## 3. The minimum coupling energy for the electrostatic "gluonic effect" in phase transition

In this section, we estimate the minimum coupling energy ($\gamma \varepsilon_2$) necessary for phase transition. The behavior of the maximum collapsed state can be understood analytically. From **Equation(4)**, we see that a maximum coupling is achieved at the symmetry breaking point $\varphi = 1/2$; thus, close to the half-occupied regime, $\varphi = 1/2$, a maximum collapsed state of the polymer solution can be expected. Because we are interested in the case of very diluted solution of multivalent salt ($c_x \to 0$), its influence can be ignored in **Equation(2)**, **Equation(5)** and **Equation(6)**. For the case of maximum coupling ($\varphi = 1/2$), fixing the volume fraction of multivalent ions at $c_x \to 0$ results in both the chemical potential $\mu$ and the osmotic pressure $\prod$ being fixed at values $\mu_0$ and $\prod_0$ respectively, which in turn mathematically casts our model to a canonical ensemble-like model for the polymer solution:

$$G - \Pi_0 = \frac{c}{N}\ln(c) + pc\ln(pc) + [1-(1+p)c]\ln[1-(1+p)c]$$
$$-\left(\ln 2 - \frac{\mu_0 + \varepsilon_1}{2}\right)pc - \frac{1}{2}\gamma\varepsilon_2(pc)^2 - \sqrt{\pi}\left(\frac{l_B}{a}pc\right)^{\frac{3}{2}} + \frac{4}{81}\left(3\pi\frac{l_B}{a}\right)^{\frac{1}{2}}pc^{\frac{3}{2}} \quad (8)$$
$$+[1-(1+p)c][p\varepsilon_{FH,1} + (1-p)\varepsilon_{FH,2}]c$$

Here the canonical-ensemble free energy (Helmholtz free energy) is given by $G - \Pi_0$.

By **Equation(8)**, we can estimate a minimum coupling energy for the "gluonic effect" that is necessary for a phase transition. This can be realized by determining the boundary condition for the spinodal decomposition of polyelectrolyte solution, which is given by $d^2(G - \Pi_0)/dc^2 = 0$ under the framework of the canonical ensemble,

$$0 = \frac{d^2(G-\Pi_0)}{dc^2} = -\gamma\varepsilon_2 p^2 - 2\varepsilon_{FH,1}(1+p)p - 2\varepsilon_{FH,2}(1-p^2) + \frac{1}{Nc} + \frac{p}{c} + \frac{(1+p)^2}{1-(1+p)c}$$
$$+\left[\frac{\sqrt{3\pi}}{27}\left(\frac{l_B}{a}\right)^{\frac{1}{2}}p - \frac{3\sqrt{\pi}}{4}\left(\frac{l_B}{a}p\right)^{\frac{3}{2}}\right]c^{-\frac{1}{2}} \quad (9)$$

Here we define an overall effective Flory parameter $2\chi_0 \equiv \gamma\varepsilon_2 p^2 + 2\varepsilon_{FH,1}(1+p)p + 2\varepsilon_{FH,2}(1-p^2)$. The minimum value of it to allow phase transition is given by $d(2\chi_0)/dc = 0$,

$$0 = \frac{d(2\chi_0)}{dc} = -\frac{1}{Nc^2} - \frac{p}{c^2} + \frac{(1+p)^3}{[1-(1+p)c]^2} + \left[\frac{3\sqrt{\pi}}{8}\left(\frac{l_B}{a}p\right)^{\frac{3}{2}} - \frac{\sqrt{3\pi}}{54}\left(\frac{l_B}{a}\right)^{\frac{1}{2}}p\right]c^{-\frac{3}{2}}. \quad (10)$$

There exists no exact explicit analytical solution for $c$ in **Equation(10)**, but a good approximation solution can be obtained by ignoring the fourth term since it is much smaller than the sum of other terms within experimental values of the parameter $l_B/a$. For the spinodal point for the collapse (away from the critical point), we obtain the approximation:

$$\frac{1}{c} \simeq (1+p) + \frac{(1+p)^{\frac{3}{2}}}{\left(\frac{1}{N} + p\right)^{\frac{1}{2}}}. \quad (11)$$

We note that the exact solution of **Equation(10)** is recovered by **Equation(11)** when the parameter $p$ approaches to zero. One can improve the analytical solution of **Equation(10)** by using the method of fixed-point iteration **[71, 72]** with the initial value defined by **Equation(11)**. However, the approximate solution of **Equation(11)** is sufficient to deduce the key features of the overall effective Flory parameter $\chi_0$ without compromising on the physical conclusions.

By insertion of **Equation(11)** into **Equation(9)**, we get an estimation for the minimum of $\chi_0$,

$$
\begin{aligned}
2\chi_{0,\min} &\equiv \gamma\varepsilon_2 p^2 + 2\varepsilon_{FH,1}(1+p)p + 2\varepsilon_{FH,2}(1-p^2) \\
&\simeq \left[\left(\frac{1}{N}+p\right)^{\frac{1}{2}}(1+p)^{\frac{1}{2}} + (1+p)\right]^2 \\
&\quad + \left[\frac{\sqrt{3\pi}}{27}\left(\frac{l_B}{a}p^2\right)^{\frac{1}{2}} - \frac{3\sqrt{\pi}}{4}\left(\frac{l_B}{a}p\right)^{\frac{3}{2}}\right]\left[(1+p) + \frac{(1+p)^{\frac{3}{2}}}{\left(\frac{1}{N}+p\right)^{\frac{1}{2}}}\right]^{\frac{1}{2}}
\end{aligned}
\quad (12)
$$

By setting $p = 0$ in **Equation(12)**, we recover the boundary condition for the spinodal decomposition of an uncharged polymer solution, i.e., $\varepsilon_{FH,2} = \frac{1}{2}\left(1 + \frac{1}{\sqrt{N}}\right)^2$. Then we get the minimum of $\gamma\varepsilon_2$ for the case of very long polyelectrolyte chain ($N \to \infty$) by **Equation(12)**,

$$
\begin{aligned}
(\gamma\varepsilon_2)_{\min} &\simeq \left[\left(\frac{1+p}{p}\right)^{\frac{1}{2}} + \frac{1+p}{p}\right]^2 - \frac{2\varepsilon_{FH,1}(1+p)p + 2\varepsilon_{FH,2}(1-p^2)}{p^2} \\
&\quad + \left[\frac{\sqrt{3\pi}}{27}\left(\frac{l_B}{a}\frac{1}{p}\right)^{\frac{1}{2}} - \frac{3\sqrt{\pi}}{4}\left(\frac{l_B}{a}\right)^{\frac{3}{2}}\right]\left[\frac{1+p}{p} + \left(\frac{1+p}{p}\right)^{\frac{3}{2}}\right]^{\frac{1}{2}}
\end{aligned}
\quad (13)
$$

A remarkable feature of the constructions of **Equation(12)** and **Equation(13)** is that: No matter the values of $\varepsilon_{FH,1} \geq 0$ and $l_B/a \geq 0$, there is an unique local maximum of $(\gamma\varepsilon_2)_{\min}$ when $\varepsilon_{FH,2}$ is larger than $\frac{1}{2}\left(1 + \frac{1}{\sqrt{N}}\right)^2$, but no local maximum of $(\gamma\varepsilon_2)_{\min}$ exists when $0 \leq \varepsilon_{FH,2} \leq \frac{1}{2}\left(1 + \frac{1}{\sqrt{N}}\right)^2$. This feature is coincidental with the well-known Θ-condition for uncharged polymers. Thus, we can already see that the necessary value of coupling energy $(\gamma\varepsilon_2)_{\min}$ to lead to a phase transition is largely regulated by the solvent quality (parameter $\varepsilon_{FH,2}$) for the uncharged part of polyelectrolyte. For most amphiphilic polymers in aqueous solutions, the value of $l_B/a$ is of order unity and $l_B/a$ ≲ 3 for vinyl polymers. As exemplars of $l_B/a = 2$ and $\varepsilon_{FH,1} = 0.45$ shown in **Figure 3** and **Figure 4**: The unique local maximum of $(\gamma\varepsilon_2)_{min}$ is regulated by the well-known poor-solvent condition for the uncharged monomers, i.e., $\varepsilon_{FH,2} > \frac{1}{2}\left(1 + \frac{1}{\sqrt{N}}\right)^2$.

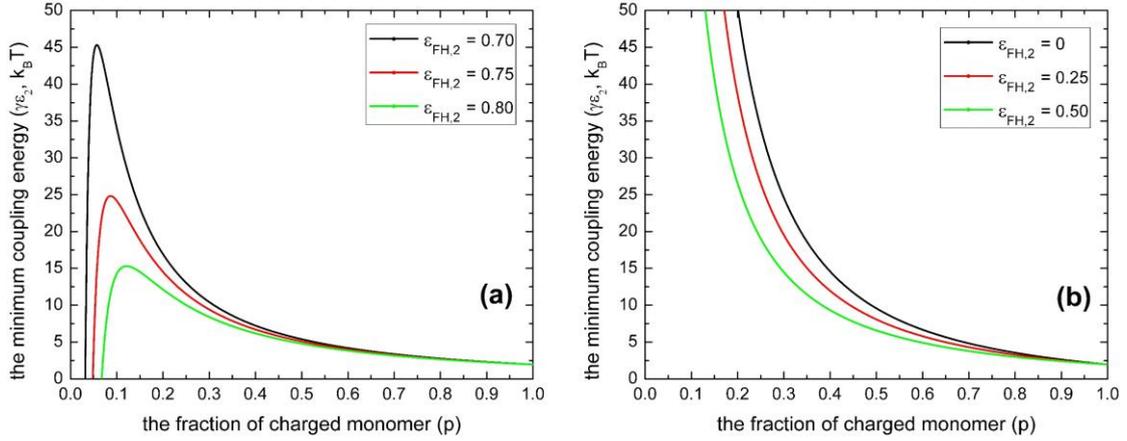

**Figure 3.** According to **Equation(13)**, the minimum coupling energy $(\gamma\varepsilon_2)$ with respect to the fraction of charged monomer ($p$) for typical values of the parameter $\varepsilon_{FH,2}$ with $l_B/a = 2$, $\varepsilon_{FH,1} = 0.45$ and $N \to \infty$: Panel **(a)** for the case of $\varepsilon_{FH,2} > \frac{1}{2}$ and panel **(b)** for the case of $0 \leq \varepsilon_{FH,2} \leq \frac{1}{2}$.

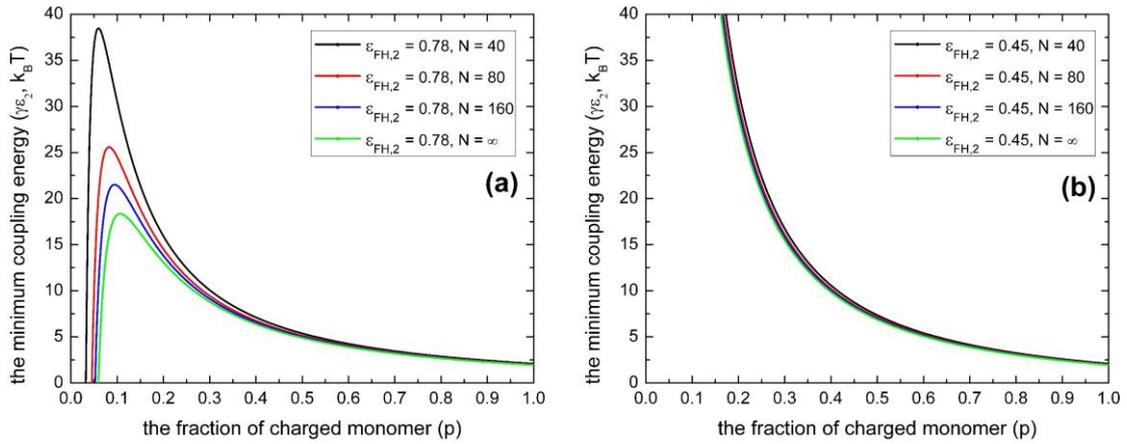

**Figure 4.** According to **Equation(12)**, the minimum coupling energy $(\gamma\varepsilon_2)$ with respect to the fraction of charged monomer ($p$) for typical polymer chain length ($N$) with $l_B/a = 2$ and $\varepsilon_{FH,1} = 0.45$: Panel **(a)** for the case of $\varepsilon_{FH,2} = 0.78$ and $\varepsilon_{FH,2} > \frac{1}{2}\left(1 + \frac{1}{\sqrt{N}}\right)^2$; panel **(b)** for the case of and $\varepsilon_{FH,2} = 0.45$ and $0 \leq \varepsilon_{FH,2} \leq \frac{1}{2}\left(1 + \frac{1}{\sqrt{N}}\right)^2$.

As indicated in **Figure 4a**, given the condition of $\varepsilon_{\text{FH},2} > \frac{1}{2}\left(1 + \frac{1}{\sqrt{N}}\right)^2$, the necessary value of coupling energy $(\gamma\varepsilon_2)_{\min}$ for a phase transition to occur can be influenced drastically by the chain length ($N$) of polyelectrolyte. In contrast, it is not sensitive to the polyelectrolyte chain length provided that $0 \leq \varepsilon_{\text{FH},2} \leq \frac{1}{2}\left(1 + \frac{1}{\sqrt{N}}\right)^2$, as shown in **Figure 4b**. This feature may be critical for understanding the re-entrant condensation of some amphiphilic proteins, since the molecular weight of proteins usually is not very large. By a comparison of **Figure 3a** and **Figure 3b**, as well as by a comparison of **Figure 4a** and **Figure 4b**, we can see that it is possible to have phase transition for polyelectrolyte when the solvent (water) quality is good for the uncharged building blocks; however, the necessary value of coupling energy $(\gamma\varepsilon_2)_{\min}$ for phase transition to occur is much higher for this situation, which is the cases reported recently by the Caltech group **[29, 30]** for poly(acrylic acid) and sodium polyacrylate in aqueous solutions of calcium salts where one calcium cation can coordinate four or more ionic acrylate monomers even though the valence of calcium cation is two. These predictions of our model are consistent with the fact that non-electrostatic interactions, such as hydrophobic interaction, play an important role in regulating the phase transition of polyelectrolyte, as indicated by recent experimental research on solution properties of proteins **[51-55]**.

## 4. The analytical phase diagrams by the model

To understand the phase behavior of polyelectrolyte under variation of concentration of multivalent ion, the free energy per monomer unit, that is, $G(\varphi, c, c_x)/c$, has to be minimized with respect to $\varphi$. Then, the replacement of the solution $\varphi(\mu, c, c_x)$ into the free energy leads to an effective free energy for the polyelectrolyte, where the effect of multivalent ions is mapped onto an effective monomer–monomer interaction which will depend on the concentration of multivalent ions. Because we are interested in the case of very negative values of $\mu$, which corresponds to the very diluted solution of multivalent salt ($c_x \to 0$), this means that in the calculation we can approximate $\varphi(\mu, c, c_x \to 0)$ by $\varphi(\mu, c)$ and approximate $G(\varphi, c, c_x \to 0)/c$ by $G(\varphi, c)/c$. Physically speaking, this approximation indicates that we focus on the adsorption and attraction effects of added small ions near/on polyelectrolyte chains, and ignore their own non-essential mixing effects if these added small ions are far away from polyelectrolyte chains. This approach will avoid heavy calculations without losing generality to get key physical conclusions.

**Equation(4)** implies that the maximum contraction of polymer chains is reached at the symmetry breaking point $\varphi = 1/2$, where the "physical crosslinking" effect reaches its maximum. This peculiar feature of the model will significantly simplify our analytical calculations. We introduce a perturbation ($\delta$) from the half occupation of the chain by the multivalent ion according to

$$\varphi = \frac{1}{2}(1-\delta). \tag{14}$$

This perturbation approach follows similarly with **refs. [57, 58, 73]**. With the expansion of $\delta$-containing terms in the logarithm function up to the accuracy of square terms ($\delta^2$) under the constraint of $|\delta| \ll 1$ and ignoring constant terms, we obtain

$$\frac{G(\delta,c)}{c} = p\left(\frac{\mu+\varepsilon_1}{2}\delta + \frac{1}{2}\delta^2\right) - \frac{1}{2}\gamma\varepsilon_2\left(1-\delta^2\right)p^2 c + \left(\frac{G_{sol}+G_{DS}+G_{FH}}{c}\right), \tag{15}$$

with

$$\frac{G_{sol}+G_{DS}+G_{FH}}{c} = \frac{\ln(c)}{N} + p\ln(pc) + \left[\frac{1}{c}-(1+p)\right]\ln\left[1-(1+p)c\right]$$

$$+\left[1-(1+p)c\right]\left[p\varepsilon_{FH,1}+(1-p)\varepsilon_{FH,2}\right] + \left[\frac{4}{81}\left(3\pi\frac{l_B}{a}\right)^{\frac{1}{2}}p - \sqrt{\pi}\left(\frac{l_B}{a}p\right)^{\frac{3}{2}}\right]\sqrt{c} + \frac{\Pi}{c}. \tag{16}$$

Minimizing the free energy in **Equation(15)** with respect to $\delta$ yields

$$\delta = -\frac{\mu+\varepsilon_1}{2(1+\gamma\varepsilon_2 pc)}. \tag{17}$$

Resubstitute **Equation(17)** into **Equation(15)**, we obtain

$$\frac{G(c)}{c} = -\frac{p}{8}\left[\frac{(\mu+\varepsilon_1)^2}{1+\gamma\varepsilon_2 pc} + 4\gamma\varepsilon_2 pc\right] + \left(\frac{G_{sol}+G_{DS}+G_{FH}}{c}\right),$$

$$= p\varepsilon_{FH,1} + (1-p)\varepsilon_{FH,2} + p\ln(p) - \frac{p}{8}(\mu+\varepsilon_1)^2 - \chi_{eff}c + g_{sol} \tag{18}$$

with the effective Flory parameter $\chi_{eff}$:

$$\chi_{eff} = \varepsilon_{FH,1}(1+p)p + \varepsilon_{FH,2}(1-p^2)$$
$$+ \frac{1}{2}\gamma\varepsilon_2 p^2 \left[1 - \frac{(\mu+\varepsilon_1)^2}{4(1+\gamma\varepsilon_2 pc)}\right] \quad , \quad (19)$$
$$+ \left[\sqrt{\pi}\left(\frac{l_B}{a}p\right)^{\frac{3}{2}} - \frac{4}{81}\left(3\pi\frac{l_B}{a}\right)^{\frac{1}{2}} p\right]\frac{1}{\sqrt{c}}$$

and the entropic term of free energy per monomer $g_{sol}$:

$$g_{sol} = \frac{\ln(c)}{N} + p\ln(c) + \left[\frac{1}{c} - (1+p)\right]\ln[1-(1+p)c] + \frac{\Pi}{c}. \quad (20)$$

In **Equation(19)**, we separated the effective Flory parameter $\chi_{eff}$ into three components. The first line of **Equation(19)** is the contribution of non-electrostatic excluded-volume interaction, the second line is the contribution of electrostatic gluonic effect between ionic monomers due to sharing multivalent salt ions, and the third line is the contribution of the non-associative electrostatic pair-wise interaction of all ions which is positive for experimental values of $l_B/a$. The effective Flory parameter $\chi_{eff}$ reaches its maximum when the value of $(\mu + \varepsilon_1)^2$ is close to zero, which corresponds to the optimally loaded state of polyelectrolyte with multivalent ions where the concentration of polyelectrolyte in condensed polymer phase reaches its maximum. This characteristic can be further observed graphically in the general features of the spinodal phase diagrams discussed in the following subsection.

The above constructed $\chi$-function is a function of the square of chemical potential $\mu$, which indicates that a $\chi$ corresponds to two values of $\mu$ and thus essentially captures the reentrant signature of polyelectrolyte condensation at lower and higher salt concentrations. In contrast, the reentrant signature of polyelectrolyte condensation cannot be well understood by the classical polyelectrolyte theories **[17-21]**, which predict that the constructed effective Flory $\chi$ parameter becomes monotonically larger as the concentration of added salt is increased, or does not change when the concentration of added salt is beyond a certain threshold value. From **Equation(19)** we point out that the reentrant signature of polyelectrolyte condensation is controlled by electrostatic gluonic effect, since only this effect is non-monotonic with respect to the salt concentration.

In order to discuss the phase transition, the pressure isotherm $\Pi(c, \mu; N, l_B/a, p, \varepsilon_1,$

$\gamma\varepsilon_2$, $\varepsilon_{FH,1}$, $\varepsilon_{FH,2}$) can be calculated from **Equation(18)** by $\partial(F/c)/\partial c = 0$, which leads to

$$\Pi = \left(\frac{1}{N}-1\right)c + \frac{p^2\gamma\varepsilon_2}{2}\left[\frac{(\mu+\varepsilon_1)^2}{4(1+\gamma\varepsilon_2 pc)^2}-1\right]c^2 - \left[\varepsilon_{FH,1}(1+p)p + \varepsilon_{FH,2}(1-p^2)\right]c^2$$
$$+ \left[\frac{2}{81}\left(3\pi\frac{l_B}{a}\right)^{\frac{1}{2}}p - \frac{\sqrt{\pi}}{2}\left(\frac{l_B}{a}p\right)^{\frac{3}{2}}\right]c^{\frac{3}{2}} - \ln\left[1-(1+p)c\right] \quad (21)$$

We note that by setting $p = 0$ in **Equation(21)** for uncharged polymer solution, with $\partial\Pi/\partial c = 0$ and the constraint of $0 < c < 1$, as it should be, we recover the boundary condition for the spinodal decomposition of uncharged polymer solution, i.e., $\varepsilon_{FH,2} > \frac{1}{2}\left(1+\frac{1}{\sqrt{N}}\right)^2$.

## 4. 1 General features of the spinodal phase diagrams

For the general case of a discontinuous phase transition, the osmotic pressure must display an unstable region of negative compressibility given by $\partial\Pi/\partial c < 0$. In **Figure 5** we display two typical examples which show the pressure isotherms for two discontinuous condensation transitions given by **Equation(21)**. The coexistence region is defined by the Maxwell construction as indicated by the horizontal isobaric lines in the figure. This cannot be obtained analytically in an exact way. However, we are able to calculate the spinodal analytically at which the solution starts to become unstable and the existence of which is the necessary condition for a discontinuous transition scenario. The spinodal of polyelectrolyte solution is given by $\partial\Pi/\partial c = 0$ and can be written in the following form:

$$(\mu+\varepsilon_1)^2 = \frac{4(1+\gamma\varepsilon_2 pc)^3}{p^2\gamma\varepsilon_2 c}\left\{\begin{array}{l}\gamma\varepsilon_2 p^2 c + \left[2\varepsilon_{FH,1}(1+p)p + 2\varepsilon_{FH,2}(1-p^2)\right]c \\ + \left[\frac{3\sqrt{\pi}}{4}\left(\frac{l_B}{a}p\right)^{\frac{3}{2}} - \frac{p}{27}\left(3\pi\frac{l_B}{a}\right)^{\frac{1}{2}}\right]\sqrt{c} \\ -\frac{1+p}{1-(1+p)c} + \left(1-\frac{1}{N}\right)\end{array}\right\} \quad (22)$$

This defines the spinodal phase diagram in the "$\mu-c$" space with the seven parameters $p$, $N$, $l_B/a$, $\varepsilon_1$, $\gamma\varepsilon_2$, $\varepsilon_{FH,1}$ and $\varepsilon_{FH,2}$. A basic characteristic of **Equation(22)** is that it predicts that the reentrant condensation of polyelectrolyte can only occur in a certain range of polymer concentration: There is no phase transition when the concentration of polyelectrolyte is too high or too low, which can be easily read from a spinodal phase diagram such as shown in **Figure 6**. It is remarkable that this prediction is consistent with experimental observations on the phase behavior of proteins in aqueous solutions of multivalent salts **[74, 75]**.

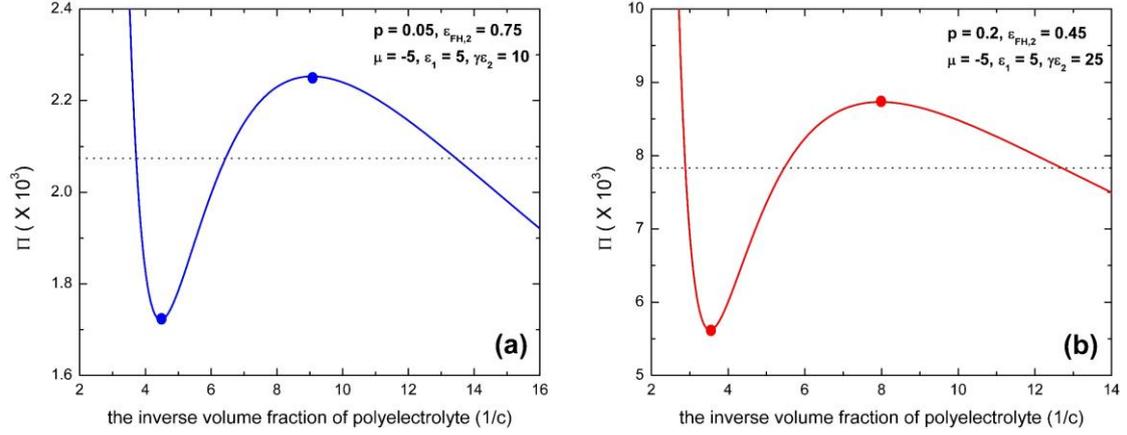

**Figure 5.** According to **Equation(21)**, osmotic pressure of polyelectrolyte solution as a function of inverse volume fraction of monomer for the case of $l_B/a$ = 2.5, $\varepsilon_{FH,1}$ = 0.45 and $N$ = 500: Panel **(a)** for the case of $p$ = 0.05, $\varepsilon_{FH,2}$ = 0.75, $\mu$ = -5, $\varepsilon_1$ = 5 and $\gamma\varepsilon_2$ = 10; panel **(b)** for the case of $p$ = 0.2, $\varepsilon_{FH,2}$ = 0.45, $\mu$ = -5, $\varepsilon_1$ = 5 and $\gamma\varepsilon_2$ = 25. The coexistence pressures by the Maxwell construction are indicated by the horizontal dotted lines in the figure, and the spinodal points are indicated by filled circles in the figure.

In **Figure 6a** we display the spinodal phase diagram by **Equation(22)** for a typical case of polyelectrolyte with parameters $p$ = 0.2, $\varepsilon_{FH,1}$ = 0.45, $\varepsilon_{FH,2}$ = 0.55, $\varepsilon_1$ = 6, $\gamma\varepsilon_2$ = 20 and $l_B/a$ = 2.5. In **Figure 6b** we display the spinodal phase diagram according to **Equation(22)** for a typical case of polyelectrolyte with parameters $p$ = 0.2, $\varepsilon_{FH,1}$ = 0.45, $\varepsilon_{FH,2}$ = 0.45, $\varepsilon_1$ = 6, $\gamma\varepsilon_2$ = 25 and $l_B/a$ = 2.5. The two solutions at the same value of $\mu$ correspond to the two extrema of the pressure isotherm, which corresponds to a coexistence of condensed and dissolved phases. The region of demixing is closed topologically. For our model, we obtain a symmetric collapse and reentry transition. The lower part of $\mu$ defines the collapse transition as indicated by the lower half of the "egg-shape" curve, while the higher part of $\mu$ defines the reentry transition. Noticeable is the strong dependence of the chain length on phase transition. A state which is condensed at a given concentration of multivalent salts can be dissolved if the chain length is reduced. However, in contrast to uncharged linear polymer **[9, 11, 57, 76, 77]**, our model predicts that it is hard to realize a real dilute phase for the reentrant condensation of polyelectrolyte when the charge fraction is not sufficiently low. As indicated in **Figure 6**, this is particularly noticeable in the limiting case of infinite chain length, there remains no small monomer concentration (*c*) in dilute phase for small values of $(\mu + \varepsilon_1)^2$ when phase separation occurs, i.e., close to the optimally loaded state of the polyelectrolyte with multivalent ions where the effective Flory parameter $\chi_{eff}$ reaches its maximum (see **Equation (19)**).

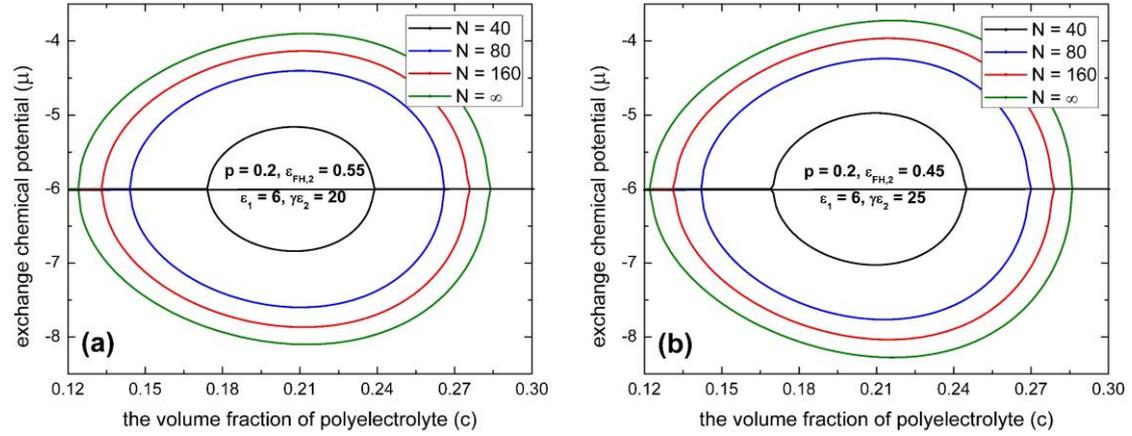

**Figure 6.** According to **Equation (22)**, spinodal phase diagrams of polyelectrolyte in the dilute solution of multivalent salts for various values of the chain lengths (*N*) with $l_B/a = 2.5$ and $\varepsilon_{FH,1} = 0.45$: Panel **(a)** for the case of $p = 0.2$, $\varepsilon_{FH,2} = 0.55$, $\varepsilon_1 = 6$ and $\gamma\varepsilon_2 = 20$; panel **(b)** for the case of $p = 0.2$, $\varepsilon_{FH,2} = 0.45$, $\varepsilon_1 = 6$ and $\gamma\varepsilon_2 = 25$.

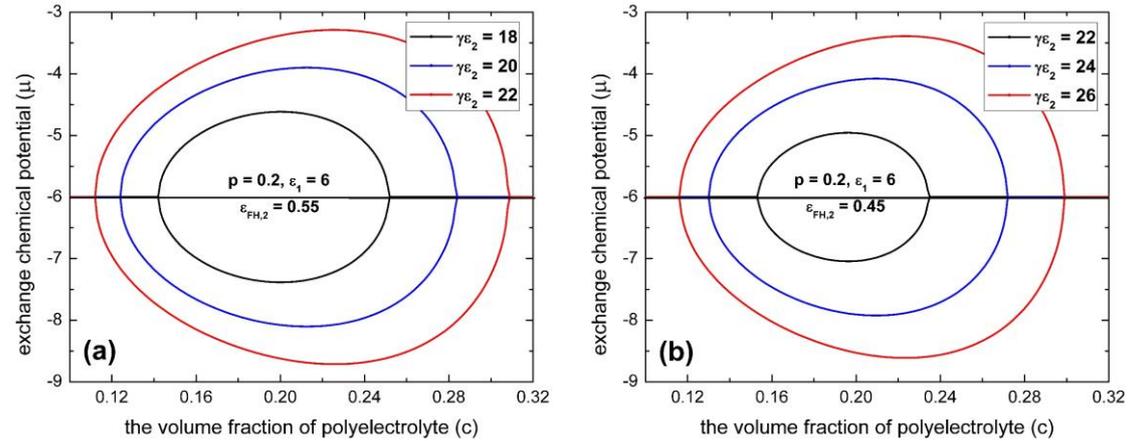

**Figure 7.** According to **Equation (22)**, spinodal phase diagrams of polyelectrolyte in the dilute solution of multivalent salts for various values of "physical crosslinking" effect ($\gamma\varepsilon_2$) with infinite chain length ($N \to \infty$), $l_B/a = 2.5$ and $\varepsilon_{FH,1} = 0.45$: Panel **(a)** for the case of $p = 0.2$, $\varepsilon_{FH,2} = 0.55$ and $\varepsilon_1 = 6$; panel **(b)** for the case of $p = 0.2$, $\varepsilon_{FH,2} = 0.45$ and $\varepsilon_1 = 6$.

Because of a significant shift of $\varepsilon_1$ on the left-hand side of **Equation (22)**, which is often not small for polyelectrolyte (on the order of about 5 $k_BT$ for the strength of an ionic bond in water at low salt concentration **[61]** and in the order of about 10 $k_BT$ for the strength of electric dipole interaction **[62]**), **Equation (22)** predicts that it melts down when the value of $\mu$ is far away from zero, which in particular corresponds to the case of very low concentrations of multivalent salts, such as shown for both the collapsed and reentry branches of the reentrant condensation of polyelectrolyte in **Figure 6**. We note that this prediction concurs with existing experimental observations on the reentrant condensation of proteins in dilute aqueous solutions of multivalent salts **[5, 8]** where the protein collapse transition occurs at rather small concentrations of multivalent salts.

In **Figure 7**, we plot the spinodal phase diagrams according to **Equation (22)** for the case of infinite chain length ($N \to \infty$) with moderate fraction of charged monomer. We see that an increase of "physical crosslinking" effect ($\gamma\varepsilon_2$) will shift the coexistence region of collapse transition to lower concentration of multivalent salts and will promote the coexistence region of reentry transition to higher concentrations of multivalent salts, which is in agreement with experimental observations for the phase transitions of both synthetic polyelectrolytes **[78-80]** and bio-polyelectrolytes **[47]** in the presence of multivalent salt. It is worth pointing out that this conclusion is also true when the fraction of charged monomer is low. However, we cannot get this conclusion by the spinodal construction of **Equation (22)**. The reason is that the osmotic pressure of spinodal calculated by **Equation (21)** can be negative when the fraction of charged monomer ($p$) is low for the case of $\varepsilon_{FH,2} > \frac{1}{2}\left(1 + \frac{1}{\sqrt{N}}\right)^2$. This scenario is thermodynamically forbidden because a stable osmotic pressure cannot be negative for polymer solutions **[9, 81]**. Now the spinodal construction of the phase diagram according to **Equation (22)** breaks down; instead, we can only numerically determine a binodal phase diagram for the case of small $p$. The details for this situation will be discussed in the next subsection.

## 4. 2 Special features of the phase diagram for the case of $\varepsilon_{FH,2} > \frac{1}{2}\left(1 + \frac{1}{\sqrt{N}}\right)^2$

For the case of $\varepsilon_{FH,2} > \frac{1}{2}\left(1 + \frac{1}{\sqrt{N}}\right)^2$, which particularly corresponding to amphiphilic polyelectrolyte, a local minimum of the osmotic pressure may exist ($\partial\Pi/\partial c = 0$) with $\Pi < 0$ for some parameter values of $p$, $N$, $l_B/a$, $\varepsilon_1$, $\gamma\varepsilon_2$, $\varepsilon_{FH,1}$ and $\varepsilon_{FH,2}$, which is particularly clear in the limiting of large $\varepsilon_{FH,2}$ with small $p$. A peculiar case is that the coexistence pressure ($\Pi$) is zero. An example of this case, shown in **Figure 8a**, indicates that by fixing other parameter values, if $p$ reaches a critical value at $p = p_0$, the corresponding Maxwell construction shows that the coexistence pressure $\Pi = 0$, and the phase transition of polyelectrolyte solution leads to a condensed polymer phase and a fluid phase without polymer. We point out that the co-existence of a condensed polymer phase and a diluted polymer phase is only possible if $p > p_0$ (for parameters used in **Figure 8**, $p_0$ is about $8 \times 10^{-4}$).

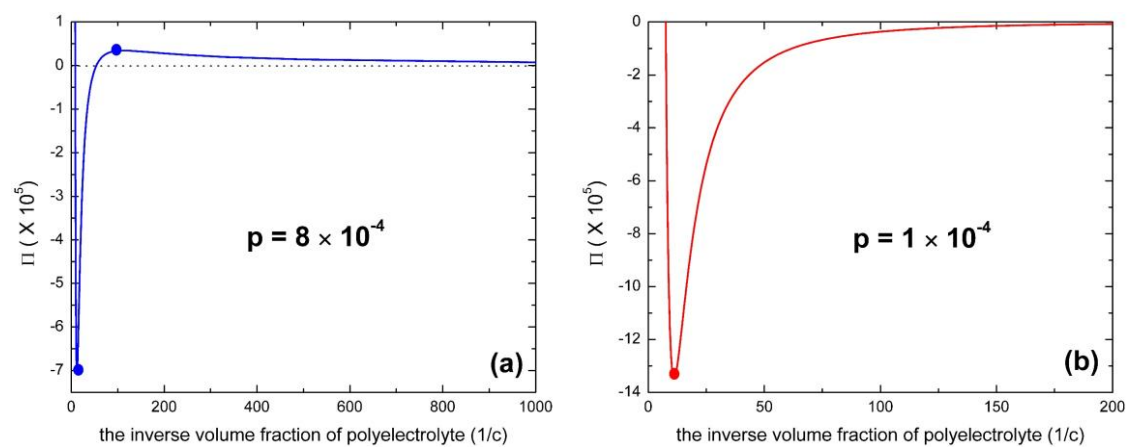

**Figure 8.** According to **Equation (21)**, osmotic pressure of polyelectrolyte solution as a function of inverse volume fraction of monomer for the case of infinite chain length ($N \to \infty$), $l_B/a = 2.5$, $\varepsilon_{FH,1} = 0.45$, $\varepsilon_{FH,2} = 0.55$, $\varepsilon_1 = 5$, $\gamma\varepsilon_2 = 15$ and $\mu = -5$: Panel **a** for the case of $p = 8 \times 10^{-4}$; panel **b** for the case of $p = 1 \times 10^{-4}$. The coexistence pressures by the Maxwell construction are indicated by the horizontal dot line in the figure, and the spinodal points are indicated by filled circles in the figure. Note that the spinodal with negative osmotic pressure in the figure is thermodynamically forbidden for polymer solutions.

When the fraction of charged monomer is less than the critical value, i.e., $p < p_0$, the corresponding Maxwell construction indicates the coexistence pressure is always negative ($\Pi < 0$), an example of this case is shown in **Figure 8b**. This scenario is thermodynamically forbidden since a stable osmotic pressure can never be negative for polymer solutions [9, 81]. For this scenario, the phase transition of polyelectrolyte solution can only result in a condensed polymer phase and a fluid phase without polymer. The polymer concentration in the condensed phase is simply given by $\Pi = 0$ instead of using the Maxwell construction and it can be written in the form of **Equation(23)**,

$$\left(\mu+\varepsilon_1\right)^2 = \frac{8\left(1+\gamma\varepsilon_2 pc\right)^2}{p^2\gamma\varepsilon_2 c}\left\{\begin{array}{l}\dfrac{p^2\gamma\varepsilon_2}{2}c+\left[\varepsilon_{FH,1}\left(1+p\right)p+\varepsilon_{FH,2}\left(1-p^2\right)\right]c \\ +\left[\dfrac{\sqrt{\pi}}{2}\left(\dfrac{l_B}{a}p\right)^{\frac{3}{2}}-\dfrac{2}{81}\left(3\pi\dfrac{l_B}{a}\right)^{\frac{1}{2}}p\right]\sqrt{c} \\ +\dfrac{\ln\left[1-\left(1+p\right)c\right]}{c}+\left(1-\dfrac{1}{N}\right)\end{array}\right\}. \quad (23)$$

**Equation (23)** has two solutions of $c$ for each value of $\mu$, i.e., $c_1 \to 0$ and $c_2 \to 1$, but only the larger one ($c_2 \to 1$) is physically correct. We find that these theoretical considerations confirm the experimental observations of the phase behavior of bovine serum albumin proteins in aqueous solutions of the multivalent salt lanthanum chloride (LaCl$_3$): At some salt concentrations, only a condensed polymer phase and no dilute polymer phase were observed in the reentrant condensation **[82]**.

By application of the Maxwell construction, to determine the parameter values at the coexistence pressure of $\Pi = 0$, we can sketch a parameter space for the co-existence of a dilute and condensed polymer phases in the phase transition of polyelectrolyte. Because the coexistence region defined by the Maxwell construction, such as shown in **Figure 8a**, cannot be obtained analytically in an exact way, we can merely get an approximation for the parameter space. Nevertheless, numerical analysis indicates that the convexity reflection points of osmotic pressure ($\Pi$) with respect to monomer concentration ($c$) is an acceptable approximation solution of $c$ for $\Pi = 0$. In other words, by a combination of **Equation(23)** with $\partial^2\Pi/\partial c^2 = 0$ for small values of $(\mu + \varepsilon_1)^2$ when phase separation occurs, i.e., close to the optimally loaded state of the polyelectrolyte with multivalent ions, we get the following approximation for the delineation line when the polyelectrolyte chain is very long ($N \to \infty$) and $p$ is small:

$$\left(\mu+\varepsilon_1\right)^2 \approx g\left(p,\gamma\varepsilon_2,\varepsilon_{FH,1},\varepsilon_{FH,2}\right) = \frac{\left[p^2\gamma\varepsilon_2 + 2\varepsilon_{FH,1}\left(1+p\right)p + 2\varepsilon_{FH,2}\left(1-p^2\right)\right]\left[p^2\gamma\varepsilon_2 + 2\varepsilon_{FH,1}\left(1+p\right)p + 2\varepsilon_{FH,2}\left(1-p^2\right) + \left(1+p\right)^2\right]}{p^2\gamma\varepsilon_2\left[\varepsilon_{FH,1}\left(1+p\right)p + \varepsilon_{FH,2}\left(1-p^2\right) + \dfrac{1}{4}\left(1+p\right)^2 - \dfrac{5}{2}p^2\gamma\varepsilon_2\right]}. \quad (24)$$

The details of deriving the above approximation can be found in **Appendix A**. To get the above approximation, for simplification, we have also ignored the influence of parameter $l_B/a$, because in the reasonable range of its experimental values, it is unimportant in our model to determine essential behaviors of the reentrant condensation of polyelectrolyte in dilute solutions of multivalent salts. The implications of this simplification will be discussed in the following subsection.

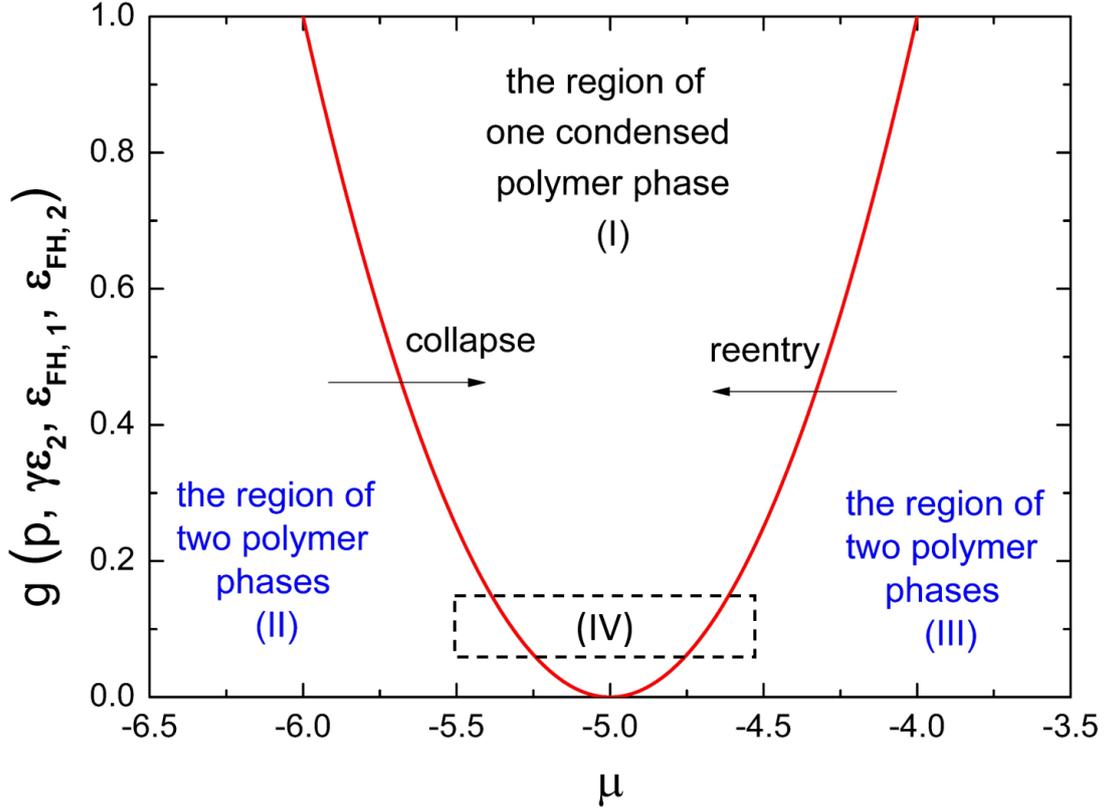

**Figure 9.** The diagram of the parameter space according to **Equation (24)** toward the co-existence of a dilute and condensed polymer phases in the phase transition of polyelectrolyte if $\varepsilon_{FH,2} > 1/2$. In the figure, the value of $\varepsilon_1$ is chosen to be 5.0. By variation of salt concentration in phase transition (such as the region IV in the figure), it is possible to see a change from the co-existence of two polymer phases to the existence of only a condensed polymer phase.

We can read some interesting physics from the **Equation (24)**. First, the enhancement of the "physical crosslinking" effect ($\gamma\varepsilon_2$) can promote the existence of only one condensed polymer phase in phase transition. This observation can be further checked by a numerical analysis of osmotic pressure even for large values of $(\mu + \varepsilon_1)^2$ when phase separation occurs, i.e., see **Equation(21)**. Second, a decrease of the incompatibility between monomers and solvent ($\varepsilon_{FH,1}$ and $\varepsilon_{FH,2}$), or an increase of the fraction of charged monomer (*p*), will promote the co-existence of a condensed polymer phase and a diluted polymer phase in the phase transition of polyelectrolyte solution. We find that this prediction is somewhat consistent with existing experimental studies on the phase transition of amphiphilic proteins **[52]**, where the concentration of protein in dilute phase becomes lower when the hydrophobic interaction between protein and water becomes stronger. Third, **Equation (24)** indicates that only a condensed polymer phase exists for uncharged polymers ($p \to 0$) if the solvent quality is poor enough, which indeed recovers the result of the classical Flory-Huggins model and can be further seen graphically from **Figure 9**.

In **Figure 9** we display the diagram of the parameter space according to **Equation (24)** toward the co-existence of a dilute and condensed polymer phases in the phase transition of polyelectrolyte. According to **Figure 9** (such as the region IV in the figure), it also becomes clear that for moderate values of $\varepsilon_{FH,2}$ and small *p* and keeping other parameters as constants, amphiphilic polyelectrolyte can show the co-existence of two polymer phases in solutions with very low or moderately low concentrations of multivalent salt. We note that this prediction concurs with the phase behavior of reentrant condensation of bovine serum albumin proteins in aqueous solutions of the multivalent salt lanthanum chloride ($LaCl_3$): In the collapse transition, only a condensed polymer phase and no dilute polymer phase were observed at some salt concentrations; while in the reentry transition, the co-existence of two polymer phases was observed at some salt concentrations **[82]**.

## 4. 3 The effect of pairwise electrostatic interaction from the correlations of all ions in reentrant condensation

The essential feature of our model for the phase separation of polyelectrolyte in the dilute solution of multivalent salts, is the induced nonlinear coupling between ionic monomers by forming "physical crosslinks" via sharing multivalent ions. An important prediction of our model is that polyelectrolyte cannot play a phase transition, merely because of the non-associative pairwise electrostatic interaction arising from the correlations of all ions in dilute solution. This can be seen clearly from **Equation(22)** by setting $\gamma \varepsilon_2 = 0$. We note that this prediction concurs with the fact that no phase separation occurs when no multivalent ions exist in dilute solutions of polyelectrolyte (e.g., see a recent excellent monograph on this topic **[9]**).

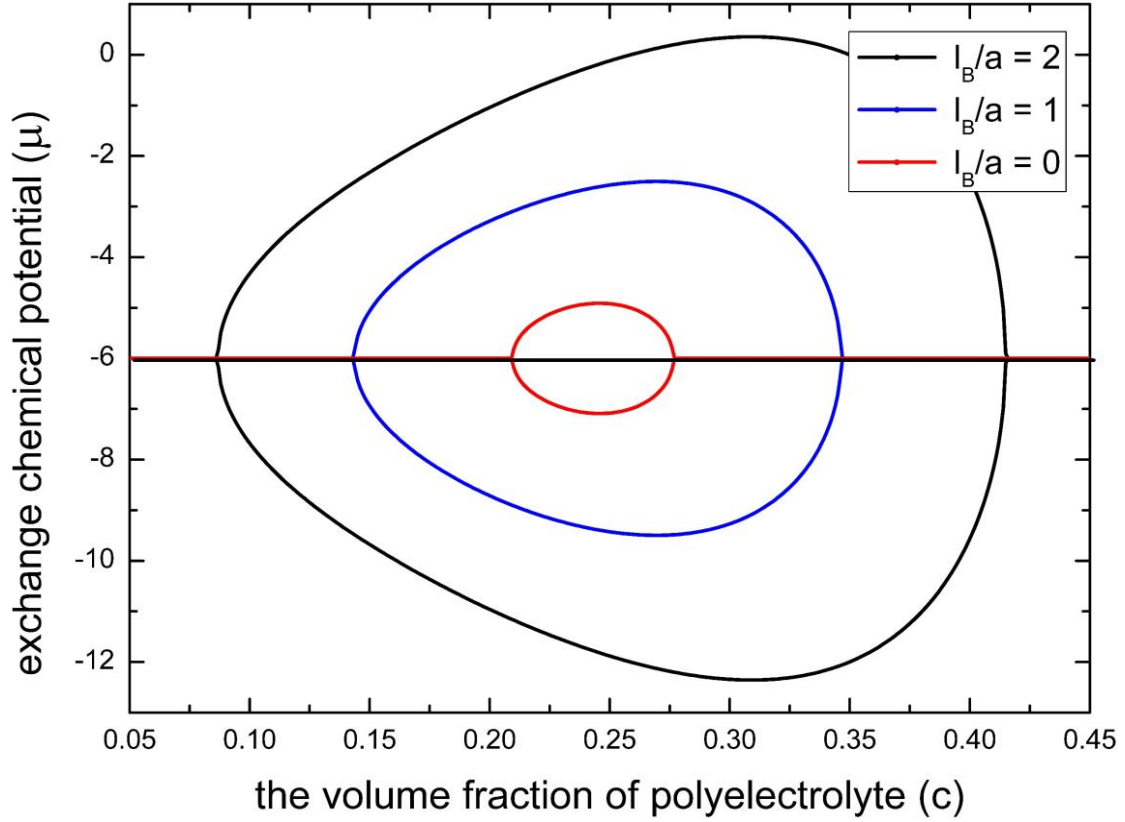

**Figure 10.** According to **Equation(22)**, by variations of the parameter $l_B/a$, an illustration of the effect of non-associative pairwise electrostatic interaction from the correlations of all ions on the phase behavior of polyelectrolyte solution. The existence of pairwise electrostatic interaction shifts the coexistence region of collapse transition to lower concentrations of multivalent salts, but shifts the coexistence region of reentry transition to higher concentrations of multivalent salts. In the figure, the parameters for the spinodal phase diagrams are chosen as infinite chain length ($N \to \infty$), $p = 0.2$, $\varepsilon_{FH,1} = 0.45$, $\varepsilon_{FH,2} = 0.72$, $\varepsilon_1 = 6$ and $\gamma\varepsilon_2 = 32$.

Nevertheless, our numerical results show that the mechanism of nonlinear coupling between ionic monomers by forming multivalent ion-assisted "physical crosslinks", can be controlled/interfered by pairwise electrostatic interaction arising from the correlations of all ions in dilute solution, which cannot be simply ignored like recent theoretical approaches **[83, 84]** for phase transition of associative polymers in which the small molecule-assisted "physical crosslinking" effect plays trivial roles. In **Figure 10**, by variations of the parameter $l_B/a$, we illustrate that the existence of pairwise electrostatic interactions can shift the coexistence region of collapse transition to lower concentrations of multivalent salts, and can shift the coexistence region of reentry transition to higher concentrations of multivalent salts. For a state close to the coexistence line, the absence of pairwise electrostatic attraction may cause a transition from the condensed to a dissolved polymer phase. It is worth noting that this effect also depends on the chain length of polyelectrolyte. We point out that these observations are manifestations of the fact that the free energy component is negative due to the electrostatic pair-wise interaction of all ions. i.e., **Equation(5).**

**4. 4 The confluence effect of non-electrostatic interactions between solvent and charged ($\varepsilon_{FH,1}$), uncharged ($\varepsilon_{FH,2}$) monomers in reentrant condensation**

The parameter $\varepsilon_{FH,1}$ in our model is not essential from a purely theoretical consideration. It can in principle be absorbed within the parameter $\varepsilon_{FH,2}$, which can be easily seen by re-defining uncharged monomers via re-grouping the uncharged moieties of polyelectrolyte together. We point out that by variations of $\varepsilon_{FH,1}$, the physical conclusions obtained in previous sections do not change. We thus avoided discussing the effect of parameter $\varepsilon_{FH,1}$ in previous sections for theoretical simplification.

However, it has practical significance for the study of amphiphilic polyelectrolyte by separating the parameter $\varepsilon_{FH,1}$ from the parameter $\varepsilon_{FH,2}$. In **Figure 11** we display the confluence effect of non-electrostatic interactions between solvent (water) and charged ($\varepsilon_{FH,1}$), uncharged ($\varepsilon_{FH,2}$) monomers in the multivalent salt-induced reentrant condensation of polyelectrolyte. We see that the confluence effect of parameters $\varepsilon_{FH,1}$ and $\varepsilon_{FH,2}$ promotes the coexistence region of collapse transition to lower concentrations of multivalent salts, and can shift the coexistence region of reentry transition to higher concentrations of multivalent salts. From the example shown in **Figure 11**, we also note that a slight poor-solvent condition for uncharged monomer ($\varepsilon_{FH,2}$ = 0.52) with a rather moderate "physical crosslinking" effect ($\gamma\varepsilon_2$ = 18) are enough for the multivalent salt-induced reentrant condensation of polyelectrolyte to occur in diluted salt solutions. This is in contrast to the cases of large values of $\gamma\varepsilon_2$ shown in previous sections, and it is common for amphiphilic polyelectrolyte such as proteins.

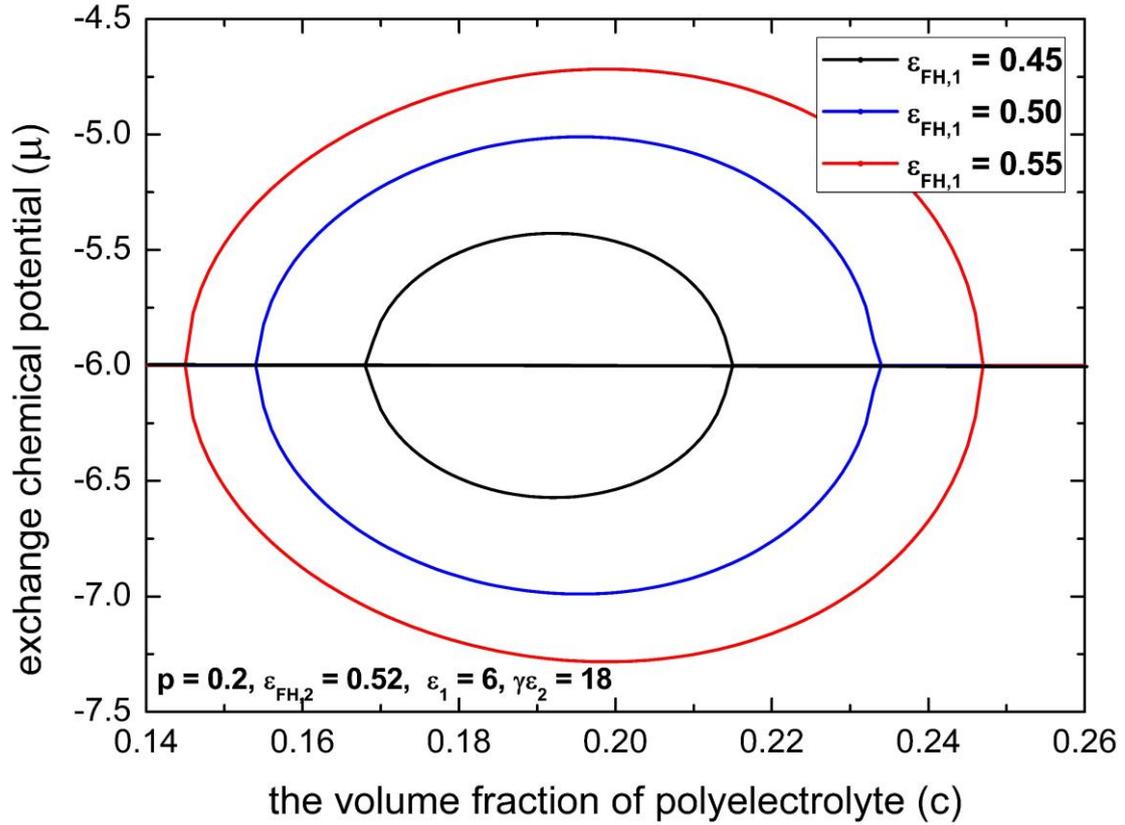

**Figure 11.** By variations of $\varepsilon_{FH,1}$ according to **Equation(22)**, an illustration of the confluence effect of non-electrostatic interactions between solvent (water) and charged ($\varepsilon_{FH,1}$), uncharged ($\varepsilon_{FH,2}$) monomers in the multivalent salt-induced reentrant condensation of polyelectrolyte. The parameters for the spinodal phase diagrams are chosen as infinite chain length ($N \to \infty$), $l_B/a$ = 2.5, $p$ = 0.2, $\varepsilon_{FH,2}$ = 0.52, $\varepsilon_1$ = 6 and $\gamma\varepsilon_2$ = 18.

## 5. Concluding remarks

The motivation of our work was focused upon the fact that polyelectrolyte cannot display phase separation merely by the non-associative pairwise electrostatic interactions, due to correlations of all ions in dilute salt solutions at room temperature **[9]**. An important consequence of analogy with the concept of cosolute-assisted "physical crosslinking" effect **[57, 58]** in this work, is that for the first time we separate the electrostatic effect into electrostatic gluonic effect due to sharing multivalent salt ions between ionic monomers, and the non-associative pair-wise electrostatic effect by correlation of all ions. This approach allows us to uncover that the electrostatic gluonic effect rather than other effects dominates the reentrant phase transition of polyelectrolyte in the dilute solutions of multivalent salts. Our theoretical calculations indicate that a minimum coupling energy for the effect of "physical crosslinking" is essential for a phase transition to occur. For the first time, this explains the puzzling experimental observation **[46-50]** that polyelectrolytes can only show a phase transition in dilute solution of salts with selective multivalency.

In contrast to uncharged linear polymer **[9, 11, 57, 76, 77]**, a distinctive finding of the present model in this work for the reentrant condensation of polyelectrolyte, is that it is hard to realize a real dilute phase when the fraction of charged monomer is not sufficiently low. This is clear from our theoretical calculations for the limiting case of infinite chain length. Our model justifies that the monomer charge is an important factor that can regulate polymer liquid-liquid phase separation. Another interesting aspect of our model in the present work is that the strong adsorption between the ionic monomer and multivalent ion can be attributed to the peculiar phenomenon that rather low concentrations of multivalent salts trigger both collapse and reentry transitions of polyelectrolyte. We have also shown theoretically that the incompatibility of the uncharged moieties of polyelectrolyte with water is critical for regulating phase behaviors of polyelectrolyte in aqueous solutions, which is in agreement with recent experimental investigations on solution properties of amphiphilic proteins **[51-55]**.

By introducing the saturated charge density $p \approx a / l_B$ with a large value of the parameter $\varepsilon_{FH,\,2} \gg 1/2$ which roughly considers the electric dipole-dipole attraction between ionized monomers if counterion condensation occurs, the corresponding spinodal phase diagrams (similar to **Figure 11** according to our model) indicate that the coexistence region of reentry transition shifts significantly to high concentrations of multivalent salts. This explains qualitatively the long-standing problem why it is easy to occur collapse transition but not reentry transition for some highly charged polyelectrolytes if without massive addition of small multivalent salts **[17, 34, 37, 79, 80]**. Nevertheless, the developed theory in this work is primarily confined to dilute solutions of small multivalent salts when the reentrant condensation of flexible polyelectrolytes occurs. This means that **Equation(5)** breaks down for medium and concentrated salt solutions even though the key formalism of **Equation(4)** is still phenomenologically applicable **[29, 30]**, and the effects of electrostatic screening or anomalous underscreening **[85]** play important roles and dipole/multipole-induced Wigner liquid phase **[65]** may form. Ways to face this challenge can be by considering the formulation of the "double screening theory" **[19, 32]** for the crossover regime **[86]** of polyelectrolyte solution, or by considering the random phase approximation **[87]** and the scaling approach **[88]** for concentrated polyelectrolyte solution. However, a detailed investigation of this aspect is worthy of future consideration and lies beyond the scope of the present study.

To conclude, motivated by recent all-atom simulation results reported by the Caltech group **[29, 30]** on phase behaviors of polyelectrolyte in aqueous solutions of multivalent salts, and by analogy with the concept of cosolute-assisted "physical crosslinking" effect proposed by Sommer **[57, 58]** to explain cononsolvency effect **[56]**, in this work we constructed a simple but effective mean-field model to explain the reentrant condensation of polyelectrolyte in dilute solutions of multivalent salts. We showed that an electrostatic gluonic effect between ionic monomers, due to sharing multivalent salt ions, plays a dominant role in governing the phase features of the reentrant condensation. This is reflected in theory by a non-monotonic concentration-dependent $\chi$-function. Moreover, we found that the interplay between electrostatic and non-electrostatic interactions, together control the phase transition of polyelectrolyte in the dilute solution of multivalent salts. This is particularly manifested by the rich solution phase behaviors of amphiphilic polyelectrolyte.

## Appendix A

By ignoring the influence of parameter $l_B/a$ and by the manipulation of $\partial^2 \Pi/\partial c^2 = 0$ for **Equation (21)**, it reads:

$$0 = \frac{\partial^2 \Pi}{\partial c^2} = -\left[\frac{(1+p)}{1-(1+p)c}\right]^2 \\ - 2\varepsilon_{FH,1}(1+p)p - 2\varepsilon_{FH,2}(1-p^2) - p^2 \gamma \varepsilon_2 \\ + \frac{p^2 \gamma \varepsilon_2}{4}(1-2\gamma \varepsilon_2 pc)\frac{(\mu+\varepsilon_1)^2}{(1+\gamma \varepsilon_2 pc)^4} \quad \text{(A1)}$$

With the expansion of $c$-containing terms up to the accuracy of linear terms under the constraint of a small fraction of charged monomers ($p$) in **Equation (A1)**, it reads:

$$c \approx \frac{\frac{p^2 \gamma \varepsilon_2}{4}(\mu+\varepsilon_1)^2 - \left[p^2 \gamma \varepsilon_2 + 2\varepsilon_{FH,1}(1+p)p + 2\varepsilon_{FH,2}(1-p^2) + (1+p)^2\right]}{\frac{3}{2}p^3(\gamma \varepsilon_2)^2 (\mu+\varepsilon_1)^2 + 2(1+p)^3} . \quad \text{(A2)}$$

With the expansion of $c$-containing terms up to the accuracy of linear terms under the constraint of small $p$ in **Equation (23)**, by an insertion of **Equation (A2)** into **Equation (23)** and ignoring the influence of parameter $l_B/a$ when chain length ($N$) is large and $p$ is small, it reads:

$$0 \approx \frac{p^4(\gamma\varepsilon_2)^2}{32}(\mu+\varepsilon_1)^4 + p^2\gamma\varepsilon_2\left[\frac{5}{4}p^2\gamma\varepsilon_2 - \frac{1}{2}\varepsilon_{FH,1}(1+p)p - \frac{1}{2}\varepsilon_{FH,2}(1-p^2) - \frac{1}{8}(1+p)^2\right](\mu+\varepsilon_1)^2$$
$$+ \frac{1}{2}\left[p^2\gamma\varepsilon_2 + 2\varepsilon_{FH,1}(1+p)p + 2\varepsilon_{FH,2}(1-p^2)\right]\left[p^2\gamma\varepsilon_2 + 2\varepsilon_{FH,1}(1+p)p + 2\varepsilon_{FH,2}(1-p^2) + (1+p)^2\right]$$
(A3)

This is a quadratic equation of $(\mu + \varepsilon_1)^2$, which can be solved exactly from model parameters $p$, $\gamma\varepsilon_2$, $\varepsilon_{FH,1}$ and $\varepsilon_{FH,2}$:

$$\frac{(\mu+\varepsilon_1)^2}{8} \approx \frac{\varepsilon_{FH,1}(1+p)p + \varepsilon_{FH,2}(1-p^2) + \frac{1}{4}(1+p)^2 - \frac{5}{2}p^2\gamma\varepsilon_2}{p^2\gamma\varepsilon_2}$$
$$+ \frac{\sqrt{6p^2\gamma\varepsilon_2\left[p^2\gamma\varepsilon_2 - \varepsilon_{FH,1}(1+p)p - \varepsilon_{FH,2}(1-p^2) - \frac{3}{2}(1+p)^2\right] + \frac{1}{16}(1+p)^4}}{p^2\gamma\varepsilon_2}.$$
(A4)

For small values of $(\mu + \varepsilon_1)^2$ when phase separation occurs, i.e., close to the optimally loaded state of the polyelectrolyte with multivalent ions, the **Equation (24)** is obtained by ignoring the first term of **Equation (A3)**. Here we quote it for convenience:

$$(\mu+\varepsilon_1)^2 \approx g(p,\gamma\varepsilon_2,\varepsilon_{FH,1},\varepsilon_{FH,2}) =$$
$$\frac{\left[p^2\gamma\varepsilon_2 + 2\varepsilon_{FH,1}(1+p)p + 2\varepsilon_{FH,2}(1-p^2)\right]\left[p^2\gamma\varepsilon_2 + 2\varepsilon_{FH,1}(1+p)p + 2\varepsilon_{FH,2}(1-p^2) + (1+p)^2\right]}{p^2\gamma\varepsilon_2\left[\varepsilon_{FH,1}(1+p)p + \varepsilon_{FH,2}(1-p^2) + \frac{1}{4}(1+p)^2 - \frac{5}{2}p^2\gamma\varepsilon_2\right]}.$$
(A5)

The **Equation (A5)** is an increasing function of parameters $\gamma\varepsilon_2$, $\varepsilon_{FH,1}$ and $\varepsilon_{FH,2}$ within their physically reasonable ranges, but it is a decreasing function of parameter $p$.

## Author Information

**Corresponding Authors**

Huaisong Yong, h.yong@utwente.nl (H. Y.)

Bilin Zhuang, bzhuang@hmc.edu (B. Z.)

Sissi de Beer, s.j.a.debeer@utwente.nl (S. de B.)

**Notes**

The authors declare no competing financial interest. This work was previously deposited as a preprint on arXiv.org at DOI: 10.48550/arXiv.2402.13686.

## Acknowledgments

Sissi de Beer and Huaisong Yong acknowledge the partial financial support for this research from the Deutsche Forschungsgemeinschaft (DFG) under the project number 422913191. Huaisong Yong acknowledges the visiting project for this research licensed by the Federal Office for Migration and Refugees of Germany (No.: BAMF 5-00002915), he also acknowledges the partial financial support for this research by the "Tianfu Emei" Scholar Foundation of Sichuan Province (No.: 2326). Huaisong Yong thank Dr. Holger Merlitz and Prof. Dr. Jens-Uwe Sommer at Leibniz-Institut für Polymerforschung Dresden, as well as Prof. Dr. Zhen-Gang Wang at Caltech for their constructive comments on the manuscript.